\documentclass{aa} 
\usepackage{csvsimple}
\usepackage{color}
\usepackage{amsmath}
\usepackage{booktabs}
\usepackage{pgfplotstable}
\usepackage{comment}
\usepackage[colorlinks=true,linkcolor=blue,citecolor=blue,filecolor=blue,urlcolor=blue]{hyperref}
\def\nic#1{\textcolor{black}{#1}} 
\def\lsim{\mathrel{\rlap{\lower3.5pt\hbox{\hskip0.5pt$\sim$}}\raise0.5pt\hbox{$<$}}}

\newcommand{\zhong}[1]{{\color{black}{#1}}}
\usepackage[labelsep=endash]{caption}

\begin{document}

\title{
Cosmology with galaxy clusters using machine learning.
Application to eROSITA Data}
\titlerunning{Machine-learning cosmological parameters by eROSITA data}

\author{Fucheng Zhong\inst{1,2,3} \and 
Nicola R. Napolitano\inst{4,1,3}\fnmsep\thanks{nicolarosario.napolitano@unina.it} \and Johan Comparat\inst{5,6} \and Klaus Dolag\inst{7,6} \and Caroline Heneka\inst{8} \and Zhiqi Huang\inst{2,9} \and Xiaodong Li\inst{2,9} \and Weipeng Lin\inst{2,9} \and Giuseppe Longo\inst{1} \and Mario Radovich\inst{10} \and Crescenzo Tortora\inst{3}}

\institute{Department of Physics E. Pancini, University Federico II, Via Cinthia 6, 80126-I, Naples, Italy \and 
School of Physics and Astronomy, Sun Yat-sen University, Zhuhai Campus, 519802, Zhuhai, P. R. China \and 
INAF – Osservatorio Astronomico di Capodimonte, Salita Moiariello 16, I-80131 Napoli, Italy \and 
Scuola Superiore Meridionale, Via Mezzocannone, 4, 80138, Naples, Italy \and
Univ. Grenoble Alpes, CNRS, Grenoble INP, LPSC-IN2P3, 53, Avenue des Martyrs, 38000, Grenoble, France \and
Max-Planck-Institut für Astrophysik, Karl-Schwarzschild-Straße 1, 85741 Garching, Germany \and
Universitäts-Sternwarte, Fakultät für Physik, Ludwig-Maximilians-Universität München, Scheinerstr.1, 81679 München, Germany \and
Institut für Theoretische Physik, Universität Heidelberg, Philosophenweg 16, 69120 Heidelberg, Germany \and
CSST Science Center for the Guangdong-Hongkong-Macau Greater Bay Area, Sun Yat-sen University, Zhuhai, 519082, China \and
INAF - Osservatorio Astronomico di Padova, via dell’Osservatorio 5, 35122 Padova, Italy
}

\abstract
{
\begin{abstract}

\textit{Context.}
Galaxy clusters provide critical constraints on cosmological parameters, e.g., through their abundance as a function of mass and redshift. However, these approaches are based on mass–observable scaling relations, which are subject to systematic uncertainties

\textit{Methods.}
We present the first Cosmological Parameter inferences using the X-ray observations of galaxy clusters by eROSITA based on a Machine Learning algorithm.  
We train a Random Forest using mock catalogs of clusters from Magneticum multi-cosmology hydrodynamical simulations. We apply the trained ML algorithm to observed X-ray features (gas luminosity, mass, and temperature) at different redshifts from the eROSITA eFEDS and eRASS1 catalogs.

\textit{Results.}
Applying the trained model to the observed cluster sample, we obtain cosmological constraints with precision comparable to those from standard analyses, such as weak lensing and cluster abundances.
We infer $\Omega_{\rm m}=0.30^{+0.03}_{-0.02}$,
$\sigma_8=0.81\pm0.01$, and $h_0=0.710\pm0.004$.
The recovered parameters show no tension in the $\Omega_{\rm m}-\sigma_8$ space, but a significant deviation of $h_0$ from the \textit{Planck} estimates. These inferences remain rather stable against variations of the input observable set and parameter space coverage. Forward modeling of the AGN feedback shows some indication that it might play a minor role. 
These results indicate that correlations among intracluster properties contain cosmological information beyond that encoded in the cluster
abundance alone, that can be captured by machine learning trained on multi-cosmology simulations.

\textit{Conclusions.}
The main result of this paper is that ML algorithms,
trained on multi-cosmology hydrodynamical simulations,
can effectively infer cosmological parameters directly from galaxy cluster data like the one provided by eROSITA. This is a change of paradigm in the context of cosmological parameter inferences.
This approach is complementary to traditional cluster count analyses and is particularly suited to large forthcoming surveys, where systematic uncertainties in mass calibration may otherwise dominate the error budget. It also highlights the potential of large-scale X-ray surveys to deliver independent tests of the standard cosmological model.

\end{abstract}

}

\maketitle
\section{Introduction}
Galaxy clusters represent the most massive gravitationally bound structures in the universe and are powerful probes of cosmology. Their abundance and large-scale distribution are sensitive to the growth of structure and the amplitude of matter fluctuations, while their thermodynamical properties encode the effects of nonlinear physics and baryonic feedback.
For that, galaxy clusters allow robust constraints on fundamental cosmological parameters such as the matter density ($\Omega_{\rm m}$), baryon fraction ($\Omega_{\rm b}$), fluctuation amplitude ($\sigma_8$), {and the dimensionless  Hubble constant ($h_0={H_0}/{100{\rm \ km \ s^{-1} \ Mpc^{-1}}}$)} that are highly complementary to both Cosmic Microwave Background (CMB) and weak-lensing (WL) probes  \citep{2011Allen_rew_complementary}. Unbiased inferences of these parameters are becoming particularly important to test the standard $\Lambda$CDM cosmological model, based on the coexistence of two dominant dark components, a Dark Energy acting as a cosmological constant ($\Lambda$) and an undisclosed form of Cold Dark Matter (CDM). Despite $\Lambda$CDM has been remarkably successful in explaining a broad range of observations  \citep{2013PhR...530...87W}, persistent tensions—such as discrepancies in measurements, between early or local universe, of the Hubble constant  \citep{2021CQGra..38o3001D, 2024ApJ...962L..17R}, the $\Omega_m -\sigma_8$ tension  \citep{2013PhR...530...87W, 2022MNRAS.509.3194P}, or the $> 2 \sigma$ anomalies in large-scale structure growth inferred from recent spectroscopic surveys \citep{2025JCAP...02..021A}—suggest either the need for modification of the cosmological paradigm or the revision
of cosmological parameter inferences, including the development of new methods, beyond
the conventional approaches.

From the theoretical side, a time-varying (dynamical) dark energy  \citep{2024JCAP...10..094C, 2024JCAP...10..048C, 2024arXiv241212905C, 2025arXiv250204212H}
or even a PhantomX coincidence  \citep{2024JCAP...12..007C} have been proposed as viable solutions for the $\sigma_8$ and the large scale structure tensions, along with voids or inhomogeneities in our local universe  \citep{2025PhRvD.111d3540G} as well as modified gravity  \citep{2024PhRvD.110l3524C, 2024SciBu..69.2698Y}, while new physics has been invoked to solve the Hubble tension  \citep{2023Univ....9...94H}.
From the methodological side, some of these tensions have been recently alleviated after having improved priors, e.g., in the lensing analysis
 \citep{2025arXiv250319441W}, thus reinforcing the line of thought that systematics in the cosmological analyses can lead to wrong parameter inferences. This highlights the necessity of developing independent and complementary approaches that can mitigate systematic uncertainties in novel ways and test cosmological models without relying on canonical assumptions. 

Here, we propose a machine learning method that uses direct X-ray measurements of galaxy clusters with minimal observational systematics. 
Traditionally, cluster cosmology relies heavily on scaling relations linking observable cluster properties, such as X-ray luminosity and temperature, to cluster mass \citep{2011ARA&A..49..409A, 2021A&A...649A.151M}. However, the mass estimates introduce significant systematic uncertainties, limiting the precision achievable with cluster-based cosmological constraints \citep{2019SSRv..215...25P, 2023A&A...675A..77R}. 
Our new approach is capable of constraining the main cosmological parameters, including $h_0$, $\Omega_{\rm m}$, and $\sigma_8$, bypassing explicit mass calibration, potentially overcoming key systematic limitations and leveraging complex, high-dimensional cluster data more effectively \citep{2015ApJ...803...50N, 2019MNRAS.484.1526A, 2019ApJ...887...25H, 2020MNRAS.499.3445Y, 2024A&A...687A...1Q}.
Motivated by the development of high-performance, large-volume, multi-cosmology simulations (e.g., Magneticum, \citealt{2016MNRAS.463.1797D}, Quijote, \citealt{2020ApJS..250....2V}), our method is based on machine learning (ML) algorithms trained on hydrodynamical simulations, characterized by different sets of cosmological parameters. We have introduced and tested this approach to effectively constrain cosmological parameters \citep{2024A&A...687A...1Q} using mock observations from \textit{Magneticum} simulations. Until now, the method has remained at the {\it proof-of-concept} stage, due to the lack of sufficiently large and uniform datasets.  

The availability of the first catalogs from eROSITA \citep{2012arXiv1209.3114M} has now filled this gap, {providing access to large, homogeneous 12,247 samples of X-ray-selected galaxy clusters with optical confirmation.} Overall, this mission is expected to detect hundreds of thousands of clusters across vast cosmic volumes \citep{2012arXiv1209.3114M, 2021A&A...647A...1P,2024A&A...682A..34M}. 
To fully exploit the potential of such an unprecedented wealth of data, {consisting of large catalogs of thousands of galaxy clusters (or features),} new analysis frameworks are needed. Using machine learning, one can more effectively correlate the cluster parameters in a multi-dimensional space, but also directly connect observational cluster features to cosmological parameters, without relying solely on traditional scaling relations. This 
can provide tighter, less biased constraints, offering novel insights into the fundamental parameters governing our Universe.

In this study, we present the first direct simulation-based inference (SBI) of cosmological parameters from real eROSITA cluster observations using a random forest machine-learning approach. 
We combine the eROSITA cluster samples with hydrodynamical simulations spanning 
a range of cosmological parameters and train a supervised Random Forrest model 
to directly infer cosmology from the multi--dimensional space of X--ray observables. 
The method uses luminosity, temperature, gas mass, and structural quantities at different redshifts as 
simultaneous inputs, thereby exploiting the internal correlations of the intracluster 
medium instead of compressing the information into a single mass proxy or scaling 
relation. In this way, the approach effectively marginalizes over the cluster mass 
calibration and reduces the dependence on external weak--lensing calibrations.

Simulation-based inference and likelihood-free methods have recently been applied to real data as alternatives to traditional likelihood analyses in cosmology \citep{2021ApJS..254...43W, 2025A&A...694A.223V, 2025MNRAS.536.1303J}. {However, 
no other methods have exploited such large-scale multi-cosmology hydrodynamical simulation suites as training sets, to produce direct posteriors on cosmological parameters.

The paper is organized as follows: in \S\ref{sec: data} we describe the observational data and 
the simulation suite used to construct the training set; in \S\ref{sec: Method} we present the inference 
method and the machine-learning framework to be applied to the eROSITA data; in \S\ref{sec: results}, we discuss the cosmological 
constraints obtained from the real cluster sample and compare the results with standard cluster cosmology analyses; in \S\ref{sec: Discussion}, we will discuss the uncertainty and systematics from the dataset and hyperparameter selection; in \S\ref{sec: conlusions}, we present our conclusions.

\section{Data description}
\label{sec: data}

The SBI analysis we propose in this work is based on two datasets: 1) the simulated galaxy cluster catalog from the multicosmology suite from {\it Magneticum}, and 2) the galaxy cluster catalog from the eROSITA observations. The former simulated galaxy cluster catalog will be used for training and testing the ML models. The observed galaxy cluster features will be used as input to constrain the cosmological parameters. In this section, we present these datasets and focus on the comparison between the \textit{feature} range and distribution, to check the domain matching between simulations and observations. Also, to fully match the observational condition of simulates clusters, we apply error-based boosting and downsampling to construct a balanced training set with uncertainties (see also \citealt{2024A&A...687A...1Q}).

\subsection{Simulation and Observation catalogs}
\label{sec: data introduction}

\pgfplotstableread[col sep=comma]{data_catalog/simulation_paras.csv}\mydata
\begin{table}
    \centering
    \caption{Cosmology parameters combination of 15 MR cosmology simulations.}
    \label{tab:mr_parameters}
    \pgfplotstabletypeset[
        string type,
        columns={name, Omega, OmegaB, Sigm8, Hubble},
        columns/name/.style={column name=Simulation, string type},
        columns/Omega/.style={column name=$\Omega_m$, string type},
        columns/OmegaB/.style={column name=$\Omega_b$, string type},
        columns/Sigm8/.style={column name=$\sigma_8$, string type},
        columns/Hubble/.style={column name=$h_0$, string type},
        column type={l},
        every head row/.style={before row=\toprule, after row=\midrule},
        every last row/.style={after row=\bottomrule},
    ]{\mydata}
\end{table}

\subsubsection{Magneticum Multicosmology Simulations} 
The simulated catalog is extracted from the middle-resolution (MR) multi-cosmology simulations \citep{2020MNRAS.494.3728S} from the Magneticum project \citep{2016MNRAS.463.1797D}, and
accessible from the Magneticum webpage\footnote{\url{http://magneticum.org/index.html}}. 
The MR simulations are ran on a box of 896 Mpc width, containing $2\times1512^3$ particles (Box1a, see Magneticum model\footnote{\url{http://magneticum.org/simulations.html}}). For the multi-cosmology suite,
15 different combinations of cosmology parameters were initially set, chosen from the vicinity of the Seven-Year Wilkinson Microwave Anisotropy Probe \citep[WMAP7]{2011ApJS..192...18K}.
The varying cosmological parameters include the matter density fraction ($\Omega_m$), baryon density fraction ($\Omega_b$), initial fluctuation amplitude ($\sigma_8$), and the Hubble constant ($h_0$), with their values listed in Table \ref{tab:mr_parameters}. 

The number of galaxy clusters in the simulated box at each redshift snapshot is shown in Fig. \ref{fig:redshift_distr}. To ensure reliable feature measurements, clusters are selected only with the halo virial mass $\text{M}_\text{vir} > 2 \times 10^{14} \boldsymbol{M}_{\odot}$. For each MR simulation, galaxy cluster features are extracted from different redshift snapshots, z=\{0.0, 0.14, 0.2, 0,47, 0.67, 0.9\}.
The extracted features of each galaxy cluster in the simulations used in this paper are summarized in Table \ref{table: features}. All features are measured within a 500 times critical density region.

\subsubsection{eROSITA catalogs} 
\label{subsec: eROSITA catalogs}
The observation catalogs are provided by the extended ROentgen Survey with an Imaging Telescope Array (eROSITA) survey   \citep{2012arXiv1209.3114M, 2021A&A...647A...1P}. One small catalog includes 542 galaxy clusters in the eROSITA Final Equatorial-Depth Survey (eFEDS,  \citealt{2022A&A...661A...7B}), and the other includes 12,247 optically confirmed galaxy clusters in the first catalog of galaxy clusters and groups in the Western Galactic Hemisphere (eRASS1,  \citealt{2024arXiv240208452B}). Those catalogs can be accessed via the eROSITA Early Data Release site\footnote{\url{https://erosita.mpe.mpg.de/edr/}} and the eROSITA-DE Data Release 1\footnote{\url{https://erosita.mpe.mpg.de/dr1/}}.
The eFEDS catalog aims to verify the performance of the Spectrum-Roentgen-Gamma (SRG)/eROSITA telescope and demonstrate its capability to detect clusters and groups at the final depth of the eROSITA all-sky survey. 
The eFEDS catalog covers a redshift range of $0.1 < z < 1.3$, with most cluster masses ranging from $10^{13} M_{\odot}$ to $3 \times 10^{14} M_{\odot}$, estimated using scaling relations (see  \citealt{2022A&A...661A...7B} for details). In comparison, the eRASS1 optically confirmed catalog covers a redshift range from $0.003$ to $1.32$ and a mass range from $5\times10^{12} M_{\odot}$ to $2\times 10^{15}M_{\odot}$.
The eFEDS catalog is estimated to have a contamination rate of $\sim 10\%$ due to high-redshift AGN contamination, while the eRASS1 catalog is estimated to have a contamination fraction $\sim 5\%$, depending on the extent of likelihood and contamination estimator threshold (see  \citealt{2024arXiv240208452B} for details).
As for the simulated features, for the observed features we also adopts the catalog quantities measured at a radius corresponding to 500 times the critical density. The features from both the simulation and observational catalogs have been converted to the same units and are listed in Table \ref{table: features}.

\begin{figure}
\vspace{-0.cm}
\hspace{-0.5cm}
    \includegraphics[width=0.51\textwidth]{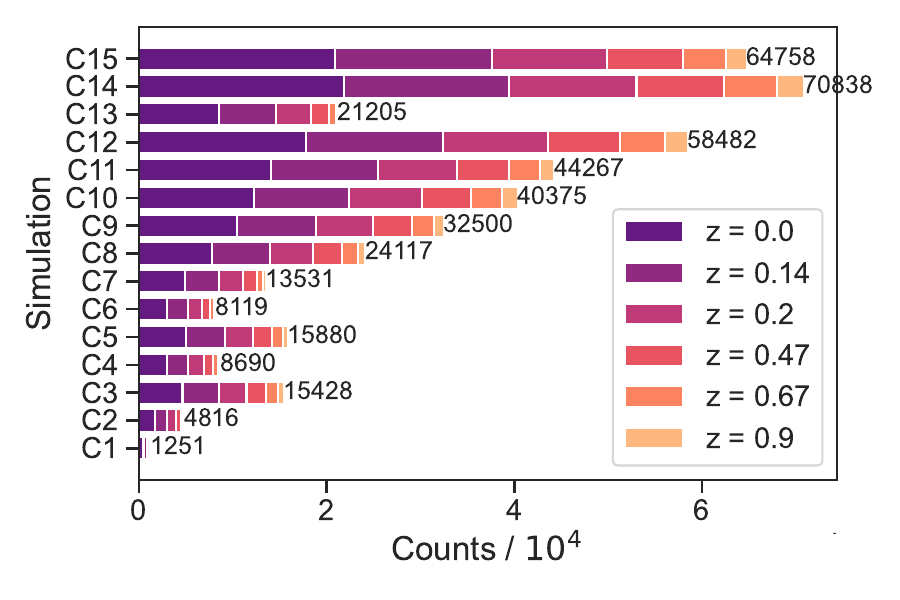}
    \caption{Galaxy cluster counts and redshift distribution of 15 MR simulations.}
    \label{fig:redshift_distr}
\end{figure}

One notable quantity is the cluster radius, $R_{500}$, which is given in units of $kpc/h_0$ or $kpc$ in the simulation catalog and eRASS1, but in units of arcmin in the eFEDS catalog.
To ensure consistency, we convert all cluster radii to physical units of $kpc$ for both simulated and observed datasets.
According to the definition in eRASS1   \citep{2024A&A...685A.106B}, $R_{500} = (\frac{3}{4 \pi} \frac{M_{500}}{\rho_c})^{1/3}$, 
where $M_{500}$ is the cluster mass within a radius enclosing a density 500 times the critical density, and $\rho_c$ is the critical density at the cluster redshift of the eROSITA fiducial cosmology.
In order 
to decouple the cosmology dependence in observations when feeding this feature to the ML classifier, 
\zhong{we have corrected all cluster radii using a dimensionless factor to preserve their units, $R = R_{500} (\frac{\rho_c}{\rho_{c,\rm fid}})^{1/3}$, where $\rho_{c,\rm fid}$ is the critical density computed under a reference fiducial flat $\Lambda$CDM cosmology with $h_0 = 0.7$ and $\Omega_m = 0.3$. The quantity $\rho_c$ is the critical density calculated under the eROSITA best-fit cosmology \citep{2024A&A...689A.298G} for each cluster. For simulated clusters, the cosmological parameters adopted in the simulation (see Table~\ref{tab:mr_parameters}) are taken as the best-fit values. This will allow us to correctly re-map the eROSITA and simulated $R_{500}$ to the same reference cosmology.}

\begin{figure}
    \centering
    \hspace{-0.cm}
    \includegraphics[width=0.4\textwidth]{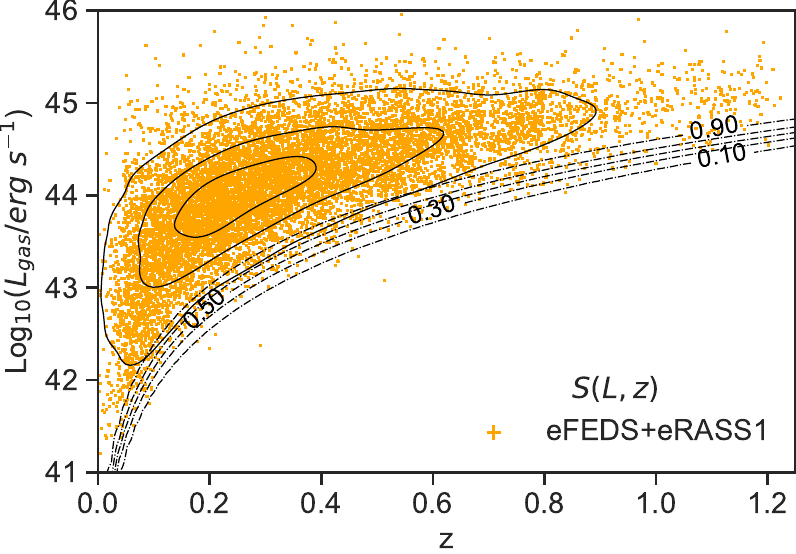}
    \caption{{Original, no cut-off, and no selection-weighted distribution of redshift and luminosity of the eROSITA cluster sample, along with the selection function used in Eq. \ref{eq: selection fun}. The probability is indicated by the dashed line, and the 16, 50, and 84 percent level contours of galaxy clusters are indicated by the solid line.}}
    \label{fig:data_L_z}
\end{figure}

To give a perspective of the cluster distribution across redshifts, in Fig. \ref{fig:data_L_z} we show the $L_{\rm gas }-z$ relation. Here we see two main features: 1) a Malmquist bias at high-$z$, where only brighter clusters are selected, and 2) a high-luminosity incompleteness at low-$z$, because of the lack of rarer massive clusters in the volume covered by eROSITA. The observational incompleteness is not incorporated in simulations; hence, to account for these effects in our cosmology inferences, in \S\ref{sec: Random-Forest constraints} we will define a selection function $S(L,z)$ to associate a probability for a cluster to be detected, to be used as a weight in our algorithm.

\begin{table}
    \centering
    \caption{Features used in this paper. The physical features are measured at the 500 times critical region (except redshift). 
    }
    \label{table: features}
    \begin{tabular}{cccccc}
        \toprule
        Feature & unit & range & description \\
        \midrule
        $\log_{10}(R_{\rm 500})$ & $kpc$ & $[2.5, \ 3.5]$ & Cluster radius\\
        $\log_{10}(M_{\rm gas})$ & $M_{\odot}$ & $[12,\ 14]$ & Gas mass\\
        $\log_{10}(L_{\rm gas})$ & $erg \cdot s^{-1}$ & $[43,\ 46]$ & X-ray bol.  lum.\\
        $\log_{10}(T_{\rm gas})$  & $keV$ & $[-0.2,\ 1.0]$ & Temperature\\
        z &  - & $[0.0, \ 1.0]$ & Redshift \\
        \bottomrule
    \end{tabular}
\end{table}

\subsection{Observational Realism: Error-based boosting and Features Normalization}

To account for typical uncertainties in X-ray features into the ML model training and prediction procedure, we define an empirical error in the training set and the observed error on observed features in the form of a Gaussian-like noise. This allows us to emulate a measurement of the observable quantities to use for training, similar to the typical errors incorporated in the observed features used for the inferences. This is a basic step to introduce ``observational realism'' to the Training set before applying this to real data (see \citealt{2024A&A...687A...1Q}, for a detailed discussion).

The simulated uncertainty is assumed to depend on the feature value measured at an overdensity around 500 times the critical density. In each cosmological simulation, the feature values are provided at overdensities of 200, 500, and 2500 times the critical density, denoted as $x_{200c}, x_{500c}$, and $x_{2500c}$, respectively. To approximate the uncertainty around $x_{500c}$, consistently with eROSITA catalogs, we adopt a simple interpolation scheme: the upper and lower uncertainties are estimated as fractional differences between neighboring overdensity measurements, as follows:
\begin{align}
    & \sigma_{+} = (x_{500c} - x_{200c})/3, \\
    & \sigma_{-} = (x_{2500c} - x_{500c})/20.
\end{align}

The boosting is implemented by by considering each cluster 
as a sample drawn from a probability distribution characterized by the actual simulated feature value and its associated errors. 

For each cluster's features, we take the logarithm (expected at the given redshift), and apply boosting using a multivariate Gaussian distribution, where the mean is given by the values of features and the standard deviation is given by the average upper and lower error, as follows:
\begin{align}\label{eq: boosting}
    & \mathcal{N}(\boldsymbol{x} | \boldsymbol{\mu}, \boldsymbol{\Sigma}) =  \frac{\exp\left( -\frac{1}{2} \ \left(\boldsymbol{x} - \boldsymbol{\mu} \right)^\top \boldsymbol{\Sigma}^{-1} \left( \boldsymbol{x} - \boldsymbol{\mu} \right) \right)}{\sqrt{(2\pi)^{d} |\boldsymbol{\Sigma}|}}, \\
    & \boldsymbol{\mu} = (\log_{10}R_{500}, \ \log_{10}M_\mathrm{gas}, \ \log_{10}L_\mathrm{gas}, \ \log_{10}T_\mathrm{gas}, \ z), \\
    & \sigma_i =  (\sigma_{i+} + \sigma_{i-})/{2}, \\
    & \Sigma_{ij} =  \sigma_i \rho_{ij} \sigma_j,
\end{align}
where $\boldsymbol{\mu}$ denotes the value of features, $\boldsymbol \sigma_+$ and $ \boldsymbol \sigma_-$ are the upper and lower errors, and $\rho_{ij}$ is the correlate coefficient between feature $i$ and $j$ in observation or simulations set. $d=5$ represents the dimension of features. For each cosmological simulation, it is randomly extracted not more than $10^4$ individual galaxy clusters, to avoid overfitting, that serve as the training samples and are subject to the above resampling procedure. For each cluster in observation and simulation, it will be resampled 100 times, then fed to the RF for training and predicting. The key hyperparameters of RF, such as the number of trees (100) and the max depth (10), were optimized via cross-validation and experimentation.

\begin{figure*}
    \hspace{-0.5cm}
    \includegraphics[width=1.03\textwidth]{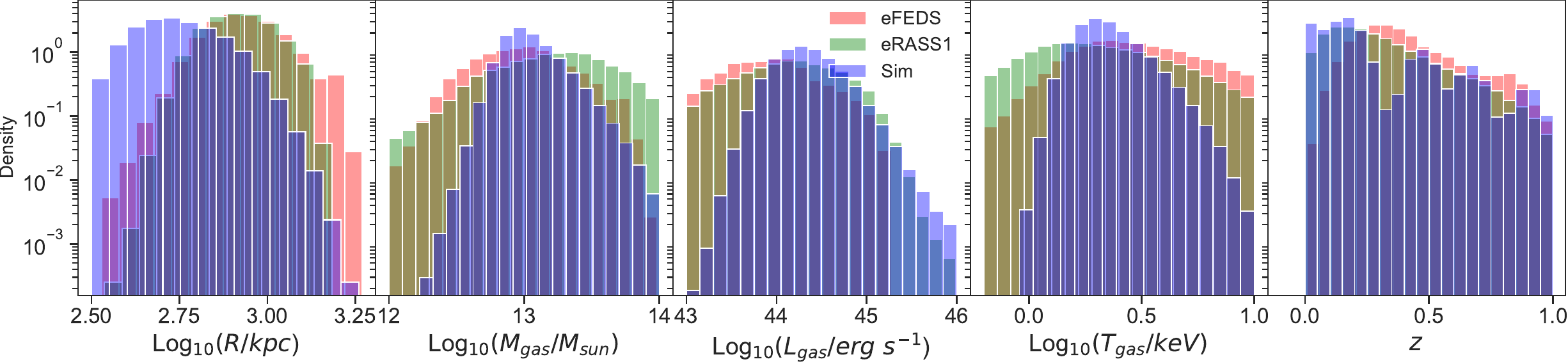}
    \caption{{Final properties of the galaxy cluster samples after cleaning: feature density distributions $(R_{500}, M_\mathrm{gas}, L_\mathrm{gas}, T_\mathrm{gas}, z)$ from the 13 Magneticum simulations (Sim) and the eROSITA observed eFEDS \citep{2022A&A...661A...7B} and eRASS1 \citep{2024A&A...685A.106B} catalogs, respectively. They are restricted to the overlapping region for effective training and inference.}}
    \label{fig:data_distr}
\end{figure*}

The boosting process is also applied to redshifts, which are given with no errors both in simulations and observation. We used the typical photometric error from photometric surveys, $\sigma_-=\sigma_+ ={0.05}$ (see, e.g.,   \citealt{2024A&A...686A.170W}), as characteristic error on redshift in simulations.



The final step is the normalization of both the training and observational datasets. Each data point in the training or observational set is normalized according to the following transformation:
\begin{align}
    F(\boldsymbol x) = \frac{\boldsymbol x - \boldsymbol x_0}{\boldsymbol x_1 - \boldsymbol x_0},
\end{align}
where $\boldsymbol x_0$ and $\boldsymbol x_1$ represent the lower and upper range of each feature, respectively, as listed in Table~\ref{table: features}. These features are equivalently mapped into the range $[0,1]$ to ensure comparable contributions to the model.


\subsection{Features distribution}
\label{sec: Features distribution and filter}

Since the range of effective inputs in a supervised ML model depends on the range of the training data, it is essential to ensure that the observed feature inputs fall within the same range. 
To achieve this, we applied the feature selection ranges listed in Table~\ref{table: features}, thereby preventing outliers from adversely impacting the results. 

Clusters with missing error-measured features in the observational data are discarded. After applying the selection filters and removing clusters with incomplete features, 346 and 2813 clusters remain in the eFEDS and eRASS1 catalogs, respectively.
These two selected catalogs are then used jointly as inputs for constraining cosmological parameters.


Regarding redshift, we decided to use only galaxy clusters in the range of 0.1 to 0.8, to avoid strong incompleteness in the sample, as shown by the $L_{\rm gas }-z$ relation in Fig. \ref{fig:data_L_z}. 
In this training-preparation phase, {among the 15 original cosmologies covered by the Magneticum simulations ($\rm C\textit{i}$, $i=1...15$) -- see Method,} we have excluded two
(C1, C2) because the $\Omega_m$ was too low to produce 
 large enough cluster number
density as compared to observations in the simulation volume for training the RF. 
After applying all these filters, the final distributions of the selected features, for both simulations and observations, are shown in the same Fig.~\ref{fig:data_distr}}. In particular, to match the observation domain, the simulated features also include the adopted errors, thus showing the actual distribution of the mock catalogs used for training and validating the ML algorithm.

\section{Method}
\label{sec: Method}
Before presenting the cosmological constraints, we describe the statistical framework used to translate cluster observables into cosmological information. Our approach relies on simulation-based inference, in which a machine-learning classifier is trained on hydrodynamical simulations sampling different cosmologies and then applied to real observations. In this section, we outline the construction and validation of the Random-Forest model, quantify its predictive performance using mock catalogs, and define how probabilistic classifications of individual clusters are combined into effective constraints on cosmological parameters.

\subsection{Random-Forest classifier}
\label{sec: Random-Forest classifier}

Random Forest \citep[RF]{ho1995random, breiman2001random} is an ensemble method that combines multiple decision trees using the bagging approach \citep{breiman1996bagging}. These trees consist of ``nodes" and ``leaves," and they are optimized by maximizing the impurity reduction or information gain (entropy reduction) as the tree grows. Each decision tree casts a vote for each label or regression value based on a given set of input features, contributing to the final prediction.

\begin{figure}
    \centering
    \includegraphics[width=0.45\textwidth]{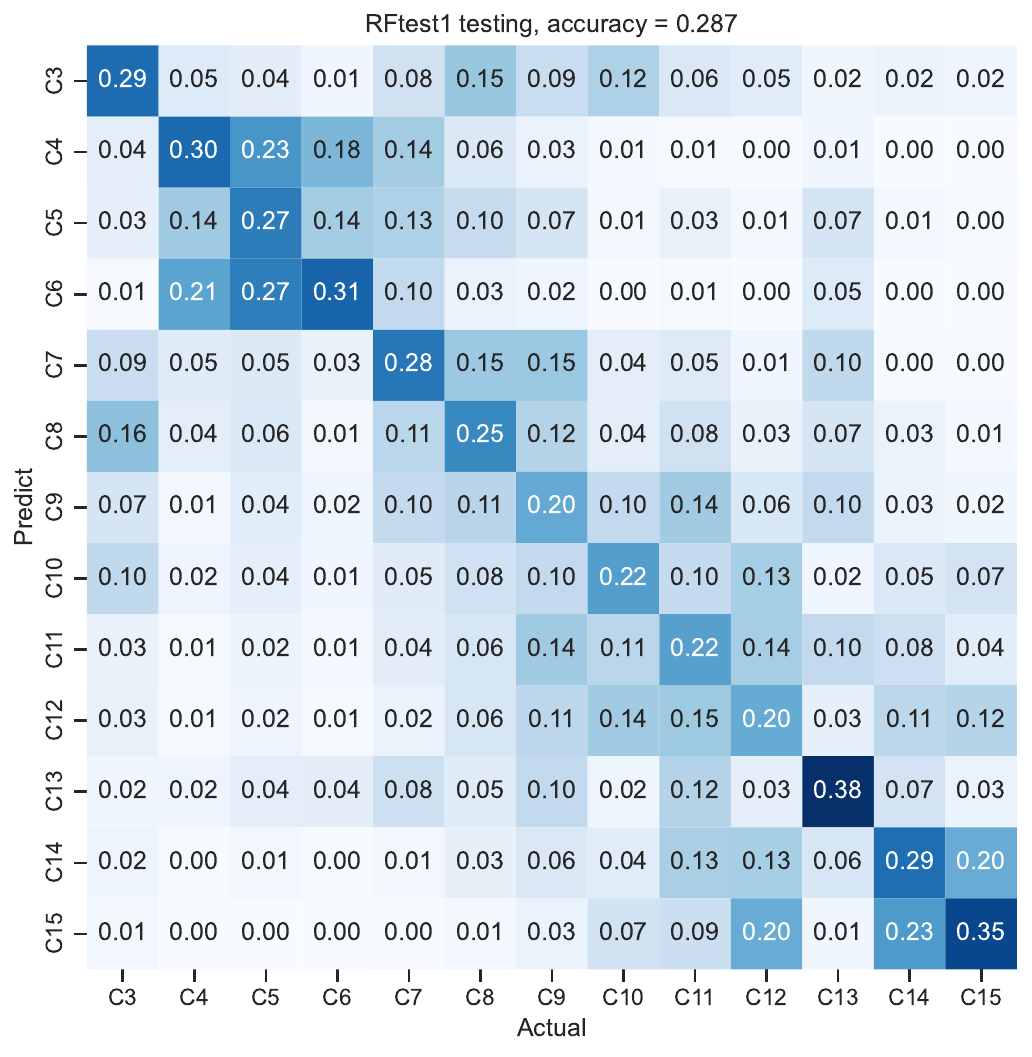}
    \caption{Confusion matrix of the $RF$ model. Along the x-axis, the reference simulations are labeled as $\rm C\textit{i}$, $i=3...15$, as C1 and C2 from the original Magneticum simulation suite are excluded because they provide too few clusters in the corresponding volume. On the y-axis, we report the classification prediction corresponding to the x-axis.}
    \label{fig: RF_baseline confusion matrix}
\end{figure}

Our RF is trained to classify catalogs of cluster observables (the features $\boldsymbol{x}$) taken from observations and determine the probability of every single cluster being drawn from one of the simulated catalogs, $P(\boldsymbol \omega | \boldsymbol x)$, corresponding to a given cosmology $\boldsymbol \omega$.
The multi-cosmology simulated catalogs from Magneticum, adopted for the training, have been re-processed to incorporate typical eROSITA observational uncertainties (see also \citealt{2024A&A...687A...1Q}) and so to generate eROSITA-like mock observations. The input observables used by the classifier include gas luminosity ($L_\mathrm{gas}$), gas mass ($M_\mathrm{gas}$), gas temperature ($T_\mathrm{gas}$) inside the characteristic radius ($R_{500}$), the $R_{500}$ itself, and the redshift ($z$).
\zhong{During the training/testing, we have used a 90\%/10\% split of the above mock eROSITA data. Moreover, by training the tuned hyperparameters across multiple trials (see Appendix \ref{sec: more tests}), we avoid overfitting and performance degradation, and obtain the fiducial model.}
After training, we assessed the accuracy of the RF using mock catalogs selected from individual simulations as the test set, and let the RF predict the cosmology behind them. In Fig.~\ref{fig: RF_baseline confusion matrix}, we derive the confusion matrix showing the overall accuracy of the method, using catalogs of 10\% mock clusters as a test set (see also \citealt{2024A&A...687A...1Q} for full tests of the algorithm). This shows that the RF generally associates a single cluster or a sample of them to a large variety of cosmologies, but the peak probability always stays on the correct one, hence giving higher ``weights'' to the correct cosmology.
\zhong{An average accuracy of 0.287 is not typical of the high accuracy observed in machine learning performances. This comes form the feature distributions across the multi-cosmology hydrodynamic simulations,
which generates a large number of similar galaxy cluster properties, despite the different cosmological parameters, thereby degrading the performance of the ML model (see also \citealt{2024A&A...687A...1Q} for a discussion). However, the principle behind our approach is not to achieve high individual galaxy cluster classification accuracy, but rather to perform statistical analysis of a large number of observed clusters. The average accuracy of 0.287 is significantly higher than the expected accuracy under random selection (1/13=0.08), indicating that the cosmological information underlying the simulations is statistically significant.}

\subsection{Constraints of cosmology parameters}
\label{sec: Random-Forest constraints}

The RF provides probabilistic predictions, assigning each observed galaxy cluster a likelihood of originating from each of the simulated cosmologies. Cosmological parameters were inferred using weighted averages, with weights derived from detection probabilities accounting for selection effects and entropy-based information gains.
When the probability of all clusters is given, the expectation of cosmological parameters is calculated by their mean, weighted by the probability of misclassifications from the confusion matrix, using the formulas:
\begin{align}
    & \boldsymbol{\bar \omega} = \sum_{ij} \boldsymbol{\omega}_i \, \mathbf{M}_{ij} \ P(\boldsymbol C_j | \boldsymbol x) / N, \\
    & N = \sum_{j} \mathbf{M}_{ij} \ P(\boldsymbol C_j | \boldsymbol x)
\end{align}
where $P(\boldsymbol C_i | \boldsymbol x)$ is the probability under the given input features $\boldsymbol x$, $\boldsymbol C_i$ is the label of one simulation, $\boldsymbol{\omega}_i$ is the corresponding values of that cosmology; $\mathbf{M}_{ij}$ is the confusion matrix shown in Fig.~\ref{fig: RF_baseline confusion matrix}, where the indices $i$ and $j$ represent the ``Actual” and ``Predicted” labels, respectively; and the $N$ is a normalized factor.
The weight of this point is calculated by the detection probabilities, $S(L,z)$, which are determined to reproduce observations (see Fig.~\ref{fig:data_L_z}) and entropy-based information gains, $\Delta H$.
The detection probability can be parameterized as a function of luminosity and redshift. {We use the flux limit, $F_{\rm lim} = 5 \times 10^{-14} \rm erg \cdot s^{-1} \cdot cm^{-2}$, corresponding to a 50\% completeness in the eROSITA survey \citep{2024A&A...682A..34M}. We use the 16th-percentile luminosity of the subsample under this flux limit, evaluated in each redshift bin, denoted as $L(z)_{\rm q16}$, as parameters of the error function ($erf$) to approximate the detection probability as follows:
\begin{align}
    & L(z)_{\rm lim} = 4 \pi  D_L(z)^2 F_{\rm lim}, \\
    & \sigma_L = L(z)_{\rm lim} - L(z)_{\rm q16}, \\ 
    & S(L,z) =  \frac{1}{2} (1 - erf(\frac{\log L(z)_{\rm lim} - \log_{10}(L)}{\sqrt{2} \sigma_L} )). \label{eq: selection fun}
\end{align}}
\nic{Note that we also consider a more conservative definition of the limiting flux, corresponding to 90\% completeness as a function of redshift. This threshold is used to define, for each object, the maximum redshift at which it would be detectable with high confidence, and hence the corresponding $V_{\max}$. Restricting the analysis to objects satisfying this criterion yields volume-limited subsamples with approximately uniform completeness across redshift.}

The variance of the predicted probabilities, $\text{Var}\left( P(\boldsymbol C_i | \boldsymbol x)\right)$, serves as an approximation of the entropy difference. The entropy difference from the maximum entropy scenario is defined as $\ \Delta H = H_{\rm max} - H$, where $H_{\rm max}$ corresponds to the case where all class probabilities are equal. $\Delta H$ represents the information gain. To avoid numerical instability caused by the logarithmic term in the entropy definition, we approximate entropy using the {\it Gini} impurity. Under this approximation, the $\Delta H$ can be estimated by the variance:
\begin{align}
    \Delta H =  n \cdot \text{Var}\left( P(\boldsymbol C_i | \boldsymbol x) \right),
\end{align}
where $n$ is the number of cosmological simulations.
For illustration, if a flat $P(\boldsymbol \omega | \boldsymbol x) = 1/n$ given by the model, the output provides no useful information, and the corresponding weight is zero.
In contrast, if the model assigns a probability of one to a single class, the variance reaches its maximum value of $\Delta H = 1 - 1/n$. The weight of each cluster is 
\begin{align}
    \Delta H /S(L,z).
\end{align}

{For each cluster, we can obtain the estimate of the cosmology parameters ($\boldsymbol{\bar \omega_j}$) and their weights ($\Delta H_j /S_j(L,z)$), with $j=1,~4$ parameters, by giving the input of their features. According to those predicted points and weights, we can calculate $ 1\sigma$ and $ 2\sigma$ contours in the $\omega_j-\omega_k$ plane with $j\neq k$, 
where we define the 
$68\%$ and $95\%$ percentile contours from the corresponding 2D Probability Density Function (PDF).} We stress, here, that the inferred parameter distributions should be interpreted as effective, likelihood-free posteriors derived from simulation-based classification, rather than exact Bayesian posteriors.

\section{Results} 
\label{sec: results}
Having established and validated the inference pipeline, we now apply the trained model to the observed eROSITA cluster catalogs. In this section, we derive the posterior distributions of the main cosmological parameters obtained from the combined eFEDS and eRASS1 samples, compare them with results from other cosmological probes, and examine the level of agreement or tension with current $\Lambda$CDM determinations.

\begin{figure}
    \centering
    \vspace{-0.1cm}
    \includegraphics[width=0.95\linewidth]{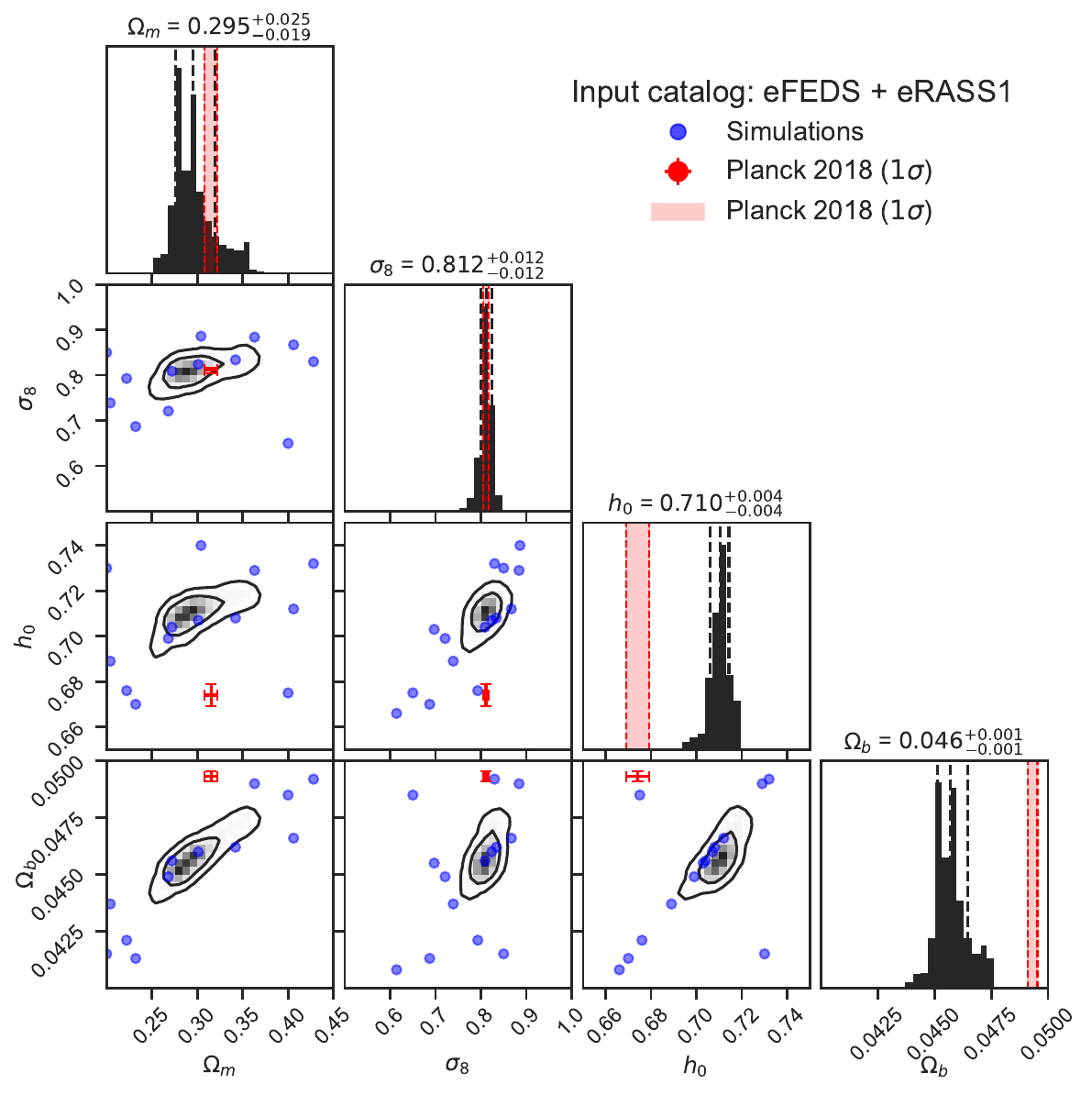}
    \caption{{Constraints on cosmological parameters $\Omega_\mathrm{m}$, $\sigma_8$, $\Omega_\mathrm{b}$, and $h_0$ derived from eROSITA clusters by the fiducial model.} 
    Dark and light shaded contours correspond to the 68\% (1$\sigma$) and 95\% (2$\sigma$) confidence intervals, respectively. Constraints from \citet{2020A&A...641A...6P} are included as red points and shaded regions, for comparison. Median values and the first and third quartiles are shown as dashed lines, and errors on the median parameter for the posterior marginalized probabilities are on top of each panel.}
    \label{fig:constraints}
\end{figure}

\begin{figure*}
    \centering
    \includegraphics[width=1.0\textwidth]{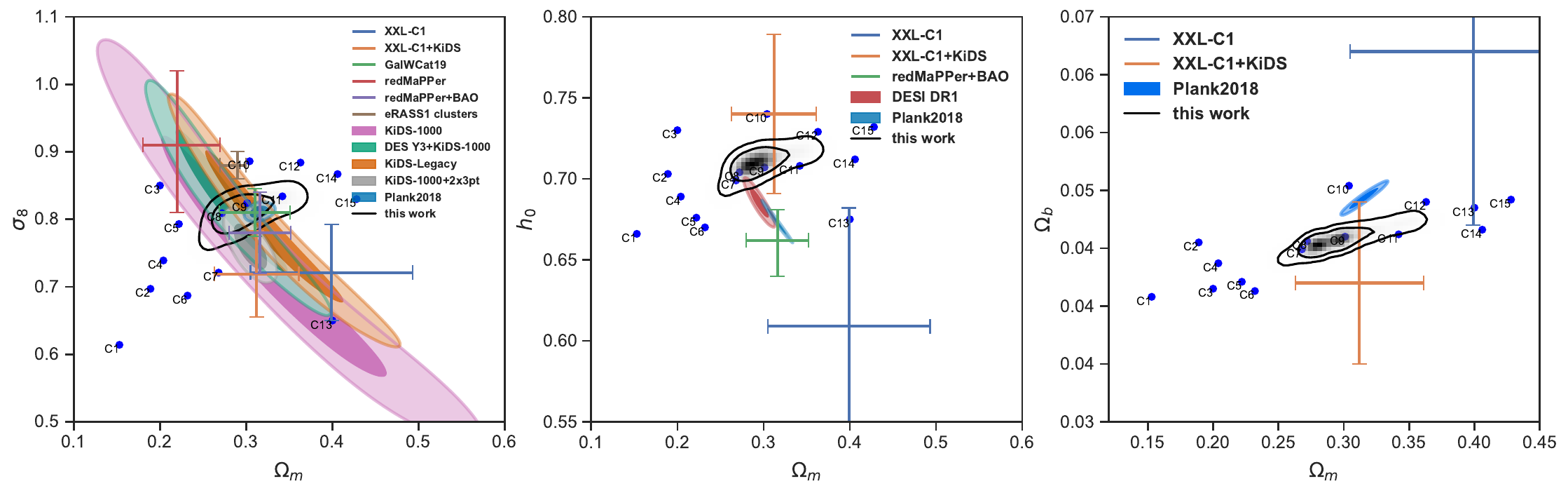}
    \caption{Constraint ($1\sigma$ and $2\sigma$ contour) of cosmology parameters compared with other literature results from other methods. These latter include XMM-XXL C1 cluster abundance, XMM-XXL C1 cluster abundance join the Kilo-Degree Survey (KiDS) tomographic weak lensing \citep{2018A&A...620A..10P}, GalWCal19 cluster abundance \citep{2020ApJ...901...90A}, SDSS RedMaPPer cluster abundance and plus BAO joint analysis \citep{2019MNRAS.488.4779C}, the eRASS1 cluster abundances \citep{2024A&A...689A.298G}, 
weak gravitational lensing of KiDS-1000 \citep{2021A&A...646A.140H}, the joint cosmic shear analysis of the Dark Energy Survey (DES Y3) and the Kilo-Degree Survey \citep{2023OJAp....6E..36D}, KiDS-Legacy cosmological constraints \citep{2025A&A...702A.169S}, KiDS-1000 combined 3×2 pt constraints \citep{2022A&A...664A.170V}, Plank cosmic microwave background \citep[CMB,][]{2020A&A...641A...6P}, and the BAO measurements from DESI DR1 data, combined with BBN and acoustic angular scale constraints \citep{2025JCAP...02..021A}.}
    \label{fig: Comparison_Constrain}
\end{figure*}

\subsection{Cosmological Inferences from Cluster Observations} 
\label{sec: cosmological parameters}
We finally use the trained RF to test real observations from eRASS1 and eFEDS and obtain direct posterior distributions of cosmological parameters, namely the matter density $\Omega_\mathrm{m}$, the amplitude of matter fluctuations $\sigma_8$, the baryon fraction $\Omega_\mathrm{b}$, and the dimensionless Hubble constant $h_0$. These are obtained by combining the weighted contribution of every cluster in the domain of the individual cosmological parameter, including the selection function $S(L,z)$ (see \S\ref{sec: Method} Methods). 
In Fig.~\ref{fig:constraints}
we plot the joint constraints of the four cosmological parameters with the related $1\sigma$ and $2\sigma$ contours, the posterior probabilities, and the median and first and third quantiles of each parameter.   
In the panels with the marginalized confidence contours, we also mark as dots the parameter values of the multi-cosmology simulations used to train the RF
as well as the ``early-universe'' parameter inferences from the Cosmic Microwave Background
by the Planck experiment \citep{2020A&A...641A...6P}, as red-shaded regions and dots with error bars. Looking at the marginalized posterior distributions, the median value of the cosmological parameters are {$\Omega_\mathrm{m}=0.30^{+0.03}_{-0.02}$, $\sigma_8=0.812^{+0.012}_{-0.012}$, 
$\Omega_\mathrm{b}=0.046^{+0.001}_{-0.001}$ and $h_0=0.710^{+0.004}_{-0.004}$,} where the quoted errors represent the 16\% and 86\% quantiles. These values are in substantial agreement (within $\sim2\sigma$) with Planck for $\Omega_m$ and $\sigma_8$, while they are discrepant for $\Omega_b$
and $h_0$, at $3\sigma$ and $>5\sigma$ level, respectively. While $\Omega_b$ discrepancy can be tracked to some systematics, as we will discuss at \S\ref{sec: Discussion},
the Hubble constant
is closer to (albeit not consistent with) the current constraints from the local universe, e.g., from Cepheids \citep{2019ApJ...876...85R}, $h_0=0.7403 \pm 0.0142 $, or SH0ES \citep{2022ApJ...934L...7R}, $h_0=0.7330 \pm 0.0104$,
and fully consistent with constraints from the Tip of the Red Giant Branch \cite[TRGB,][]{2019ApJ...882...34F}.
Taking these results at face value, we can conclude that there is no $\Omega_m-\sigma_8$ tension with Planck, while our simulation-based cluster inference of $h_0$ mirrors the TRGB-based estimates, effectively bridging the constraints between early and late Universe probes.

Because of the sparse and nonuniform coverage of the parameter space (see also \S\ref{sec:param_cover}), we have checked if the shape of the confidence contours is affected by the specific configuration of the parameter priors. Looking at the Fig. \ref{fig:constraints}, this is not the case since, for instance, the $h_0-\Omega_b$ contours strongly deviate from the aligned distribution of priors.  To further check that the contours are not fictitiously elongated in the $\Omega_m-\sigma_8$ and  $\Omega_m-h_0$, 
we have produced a series of tests where we have excluded cosmologies at $\Omega_m>0.35$ or $h_0>0.72$ and found that the contours and the 
parameters are weakly affected and remain consistent with the fiducial estimates (see Appendix \ref{sec: more tests}), unless these latter are also excluded from the training set. In this latter case, the RF still produces parameter constraints but with a lower degree of confidence.

\subsection{RF inferences in the Cosmological Context}
\label{sec: cosmological context}

We can finally put the inferences obtained in the previous section into a cosmological context. In Fig.~\ref{fig: Comparison_Constrain} we compare our results on $\Omega_{m}$ vs. $\sigma_8$, $h_0$ and $\Omega_b$ with those from the Planck CMB observations \citep{2020A&A...641A...6P}, galaxy clustering and lensing from the Dark Energy Survey \citep[DES]{2022PhRvD.105b3520A}, recent BAO results from DESI \citep{2025JCAP...02..021A}, and other cluster-based cosmology analyses \citep{2018A&A...620A..10P,2019ApJ...878...55B,2022MNRAS.510..131M,2024A&A...689A.298G}. From this figure, we see that there is some consistency across these independent cosmological probes, but also evident discrepancies. Starting with the $\Omega_m-\sigma_8$ plot, our ML contours lie orthogonally to the WL constraints and overlap at less than 1$\sigma$ level with the Planck CMB constraints, i.e., showing no tension with the CMB results. Note that the shape of the contours is only partially dictated by the ``prior'' distribution represented by the simulation parameters marked as dots in the same figure. This can be seen in Fig.~\ref{fig:constraints} where, e.g., in the $\sigma_8$ and $h_0$ columns, the contours do not mirror the priors, but it has also been confirmed by changing the prior distribution using different subsets of cosmologies (see Appendix \ref{sec: more tests}). However, expanding the parameter space of the multi-cosmology suites is a crucial step toward fully exploring potential degeneracies among the cosmological parameters. For the time being, our ``Fiducial'' $\Omega_m-\sigma_8$ constraints are consistent with the latest results of weak lensing showing no tension with Planck (e.g., KiDS,  \citealt{kids_legacy_WL}). On the other hand, looking at the middle panel of Fig.~\ref{fig: Comparison_Constrain}, we see a significant tension on the Hubble constant with the CMB value, consistent with the XXL cluster survey when combined with weak lensing constraints from KiDS \citep{2018A&A...620A..10P}.
Finally, in the right panel of the Figure, 
our analysis reveals a $>2\sigma$ tension in the $\Omega_b-\Omega_m$ space with the 
Planck constraints.  
The origin of this offset may be partially tracked to boundary effects arising from the finite support of the sampled parameter space. Indeed in Fig. \ref{fig: Comparison_Constrain}, we can see that the Planck contraints fall at the edge of the Magneticum parameter space and this cause the RF to be less sensitive to these cosmologies. While fully addressing boundary-related effects ultimately requires simulations spanning a wider cosmological parameter space, in \S\ref{sec: Discussion} we will try to quantify the possible impact of this precondition. On the other hand, this ``tension'' may also reflect a genuine under-abundance of baryons in galaxy clusters relative to CMB-based inferences, as reported in previous studies using standard gas-fraction analyses \citep{Eckert2019, Wicker_2023A&A...674A..48W}, or be the consequence of systematic gas depletion driven by AGN feedback \citep{Cusworth2014}.
Because our simulation suite does not include multiple feedback prescriptions (with the exception of the C8 cosmology), to assess whether variations in feedback strength could shift the inferred $\Omega_b$ values toward Planck-compliant cosmologies, we attempt to mimic the impact of stronger or weaker feedback models. This is tested in Appendix \ref{sec: agn feedback} where we quantify the influence of these ``altered'' feedback recipes on the inferred $\Omega_b$ and other cosmological parameters. 
There, we find that feedback variations play only a minor role, suggesting that the observed $\Omega_b$ tension is more closely related to the completeness of the parameter-space coverage than to uncertainties in the feedback modeling.

\section{Discussion}
\label{sec: Discussion}
The primary result of this paper is that ML algorithms, trained on multi-cosmology hydrodynamical simulations, can effectively infer cosmological parameters directly from galaxy cluster data like the one provided by eROSITA. This is a change of paradigm in the context of cosmological parameter inferences, as the possibility to exploit the cosmological information in multi-cosmology hydro-simulations to constrain cosmology was theorized \citep{2023ApJS..265...54V,2024A&A...687A...1Q}, but not yet applied to real data. The great advantage of our simulation-based method, applied to galaxy clusters, is that it provides an orthogonal approach to cosmology, significantly mitigating the systematic uncertainties associated with traditional mass calibrations \citep{2011ARA&A..49..409A, Kravtsov_Borgani_2012, 2019SSRv..215...25P, 2022A&A...665A.100L, 2022MNRAS.511.1484I}. On the contrary, it relies entirely on direct, X-ray-measured quantities, with almost no cosmological assumptions except for $R_{500}$, which we have renormalised to account for its cosmological dependence when used as an input feature (\S\ref{subsec: eROSITA catalogs}).   
In the previous section, we have seen that, by directly translating cluster observables into cosmological constraints, our approach achieves precision and accuracy comparable to current state-of-the-art cosmological methods from WL, large-scale structure, cluster mass function, and CMB measurements.
Our inferences, though, stand on a series of ingredients that may suffer from systematics. In this section, we want to fully address all possible sources of systematic errors to assess the robustness of our results.

\subsection{Selection function}
An accurate characterization of the survey selection function is essential for unbiased cosmological inference from flux-limited cluster samples. 
In our analysis, the selection function $S(L,z)$ encodes the detection probability of a cluster as a function of its X-ray luminosity and redshift, and it is used as a weight during inference to account for the variable completeness. Simulations also suffer incompleteness: on the high mass end, very massive clusters can be lost depending on the volume of the simulations, similarly to the observations, while at the low mass end, it is the mass resolution that defines the minimum agglomerate of particles recognisable as a bound system. These incompleteness, which are redshift dependent, impact in different ways the \textit{training} and the \textit{test} samples, and this can produce systematics in the cosmological parameter inferences.  

\begin{figure}
    \hspace{-0.4cm}
    \includegraphics[width=0.51\textwidth]{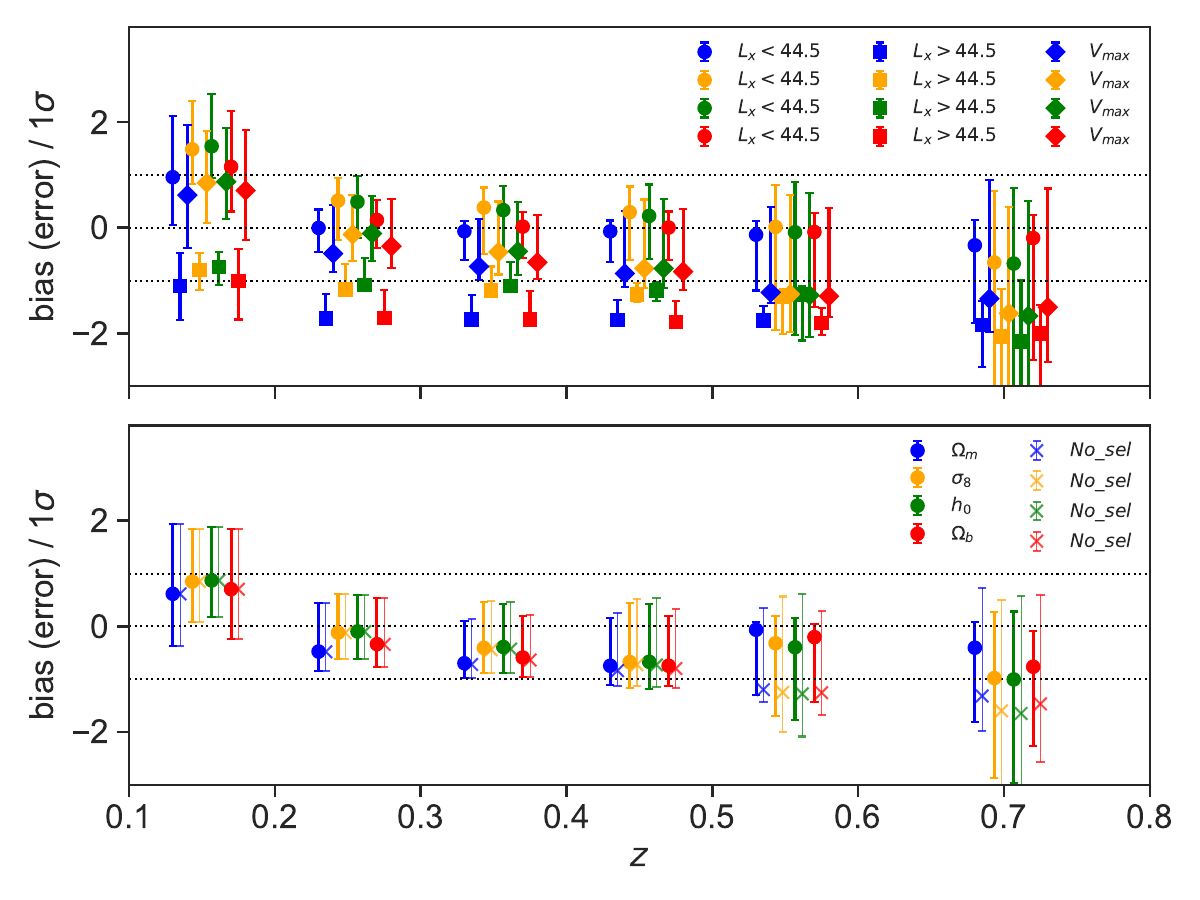}
    \caption{Normalised bias of the cosmological inferences as a function of redshift. 
    The y-axis represents the deviation of the cosmological parameters with respect to the fiducial model inference (bias) 
    normalized by the 1$\sigma$ error. The upper panel shows the $\text{bias}/1\sigma$ for the \textit{bright} and \textit{faint} samples (see text) using no selection function, {and the total samples within a $V_{\rm max}$ volume.}
   The bottom panel shows the $\text{bias}/1\sigma$ for the whole cluster sample, either implementing or not the selection function. From this figure, we see that the implementation of the $S(L,z)$ produces inferences from the eROSITA sample with no systematic trends across the redshift bins and alleviates the impact of the bias induced by the \textit{bright sample} if no selection function is applied.} 
    \label{fig:Parameters_vs_z}
\end{figure}

To check this, we have performed a controlled experiment in which the {eROSITA} cluster sample was divided into two luminosity ranges: 
a {\it bright subsample} ($\log L_{\rm gas} > 44.5$ erg $\rm s^{-1}$), expected to be nearly complete across the full redshift range (except possibly at $z<0.2$), and a {\it faint subsample} ($\log L_{\rm gas} < 44.5$ erg $\rm s^{-1}$), where incompleteness effects are pronounced at $z>0.3-0.4$. \nic{Correspondingly, we have also selected a volume-limited sample, i.e., by selecting only clusters brighter than the limiting flux corresponding to the 90\% detection probability ($V_{\rm max}$) as a function of redshift (see also \S\ref{sec: Method}).}
For each subsample, we repeated the full machine learning inference and compared the resulting cosmological parameters with those obtained from the fiducial model, which includes the full sample weighted by $S(L,z)$. The results are shown in Fig.~\ref{fig:Parameters_vs_z}, where we plot the deviation of all parameters from fiducial inferences (bias) of Fig.~\ref{fig:constraints} as a function of the redshift, normalised by the overall 1$\sigma$
error in the full range of redshift, $\text{bias}/1\sigma$. This is defined as $(E - \bar E)/\bar \sigma_{\text{sign}(E - \bar E))}$, where $E$ is the expectation value and $\text{sign}$ denotes the sign function. 
In the top panel, the \textit{faint} subsample yields cosmological constraints fully consistent with fiducial inferences, except for the first redshift bin. In contrast, the \textit{bright} subsample
produces a negative bias in the parameters, except eventually for the first bin. 
Despite this may look counterintuitive, as one would expect the more massive sample to carry more accurate cosmological information, it is possibly the consequence of a stronger effect of the high mass incompleteness, mimicking a lower mass density and $\sigma_8$ cosmologies, with respect to the low mass incompleteness. \nic{Interestingly, the ``$V_{\rm max}$ sample'' inferences tend to align with the  \textit{faint} subsample inferences at low-$z$, and to the \textit{bright} subsample inferences at high-$z$, due to the shift of the completeness of observations toward brighter fluxes at high-$z$.} To confirm that this is an effect driven by data incompleteness and not by the RF model,
we have performed a similar test on mock catalogs from different simulations, and none of them showed any systematic bias.
This is diluted once $S(L,~z)$ is applied, as seen in the bottom panel of Fig.~\ref{fig:Parameters_vs_z}. Here, despite a tendency to a negative bias of the inferences in the redshift bins, there is no systematic bias if the correction via the $S(L,~z)$ is applied, while a more systematic trend is seen if the effect of the $S(L,~z)$ is neglected (``no sel'' models). Overall, this test demonstrates that our fiducial weighting scheme based on $S(L,z)$ effectively mitigates incompleteness-driven biases and ensures that the cosmological constraints are equally driven by all clusters in the sample giving more weight to the incomplete low-mass sample. Future data releases with larger cluster samples and higher-resolution cosmological simulations will be crucial to test further the effects of incompleteness. 

\begin{figure}
    \hspace{-0.8cm}
    \includegraphics[width=0.54\textwidth]{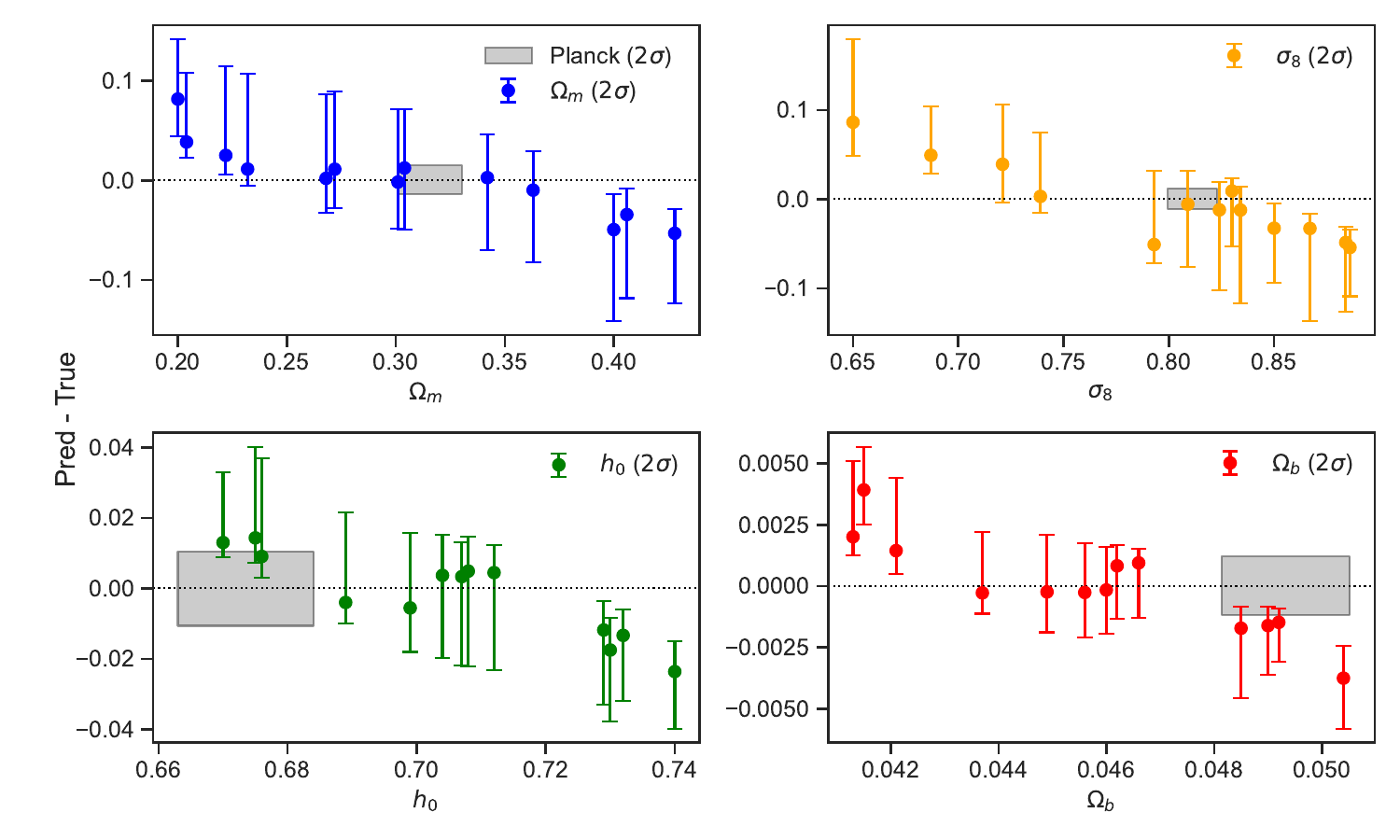}
    \caption{Finite-support effects in the simulation-trained inference. Difference between inferred and true cosmological parameters as a function of the input values, measured from mock cluster catalogues. Grey bands indicate the Planck 2018 (2$\sigma$) constraints. The trends reveal mild boundary effects near the edges of the training parameter space, most evident for $h_0$ and $\Omega_{\rm b}$. However, the magnitude of these effects is insufficient to account for the tensions observed in the real-data inference.}
    \label{fig:boundary_test}
\end{figure}

\subsection{Finite parameter-space coverage} 
\label{sec:param_cover}
\nic{Simulation-based inference methods trained on a discrete and finite set of cosmological models may exhibit support-mismatch and boundary effects when the true parameters lie close to the edges of the sampled domain. 
To quantify the impact of the support-mismatch and boundary effects on our analysis, we perform end-to-end validation tests using mock catalogs constructed from individual cosmologies in the simulation suite, following the same procedure adopted for the real eROSITA data.
For each cosmology, we generate synthetic cluster catalogs and apply the full inference pipeline, including feature extraction, classification, and weighting. We then compute the discrepancy between the inferred and true cosmological parameters and examine its dependence on the input parameter values.
This is shown in Fig.~\ref{fig:boundary_test}, where can see that some biases can indeed emerge near the edges of the sampled parameter space, where the systematics can reach no more that 10-15\%. 
This effect is noticeable for $\Omega_b$ and for $h_0$, particularly because the Planck 2018 reference values lie close to the boundary of the sampled parameter space for these two parameters. This raises the possibility that part of the observed offsets observed in Fig. \ref{fig: Comparison_Constrain} could be artificially induced by edge effects. However, the magnitude of the bias inferred from the mock-catalog tests in the vicinity of the Planck values is of the order of $2\sigma$, which is significantly smaller than the deviations observed for $h_0$, but comparable to the offset measured for $\Omega_b$ (Fig.~\ref{fig: Comparison_Constrain}). This leads us to conclude that the reported $h_0$ tension cannot be dominated by finite-support effects, while $\Omega_b$ possibly can (see below).}
Nevertheless, these results highlight the importance of extending multi-cosmology simulation suites, particularly along the $\Omega_b$ and $h_0$ directions, to further reduce boundary-related systematics in future applications of the method.

\begin{figure}
    \centering
    \includegraphics[width=0.45\textwidth]{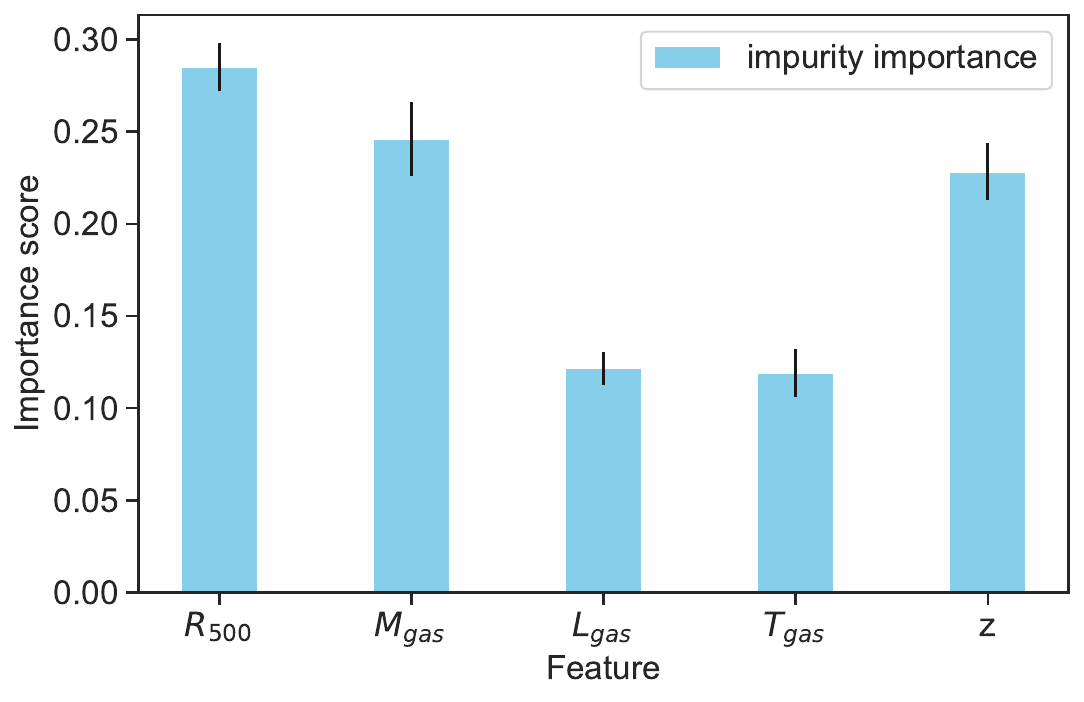}
    \caption{Feature importance from random forest classifiers. The impurity-based feature importance is reported, indicating that the relative importance indicates radius ($R_{500}$) and gas mass ($M_\mathrm{gas}$), together with redshift, dominate cosmological inference.}
    \label{fig:feature}
\end{figure}

\subsection{The impact of the $R_{500}$}
In \S\ref{subsec: eROSITA catalogs} we have seen that  $R_{500}$ is a cosmology-dependent quantity 
and discussed the correction adopted for the $R_{500}$ from eROSITA catalogs to make them independent of the cosmology. This correction is applied directly to the radius  used as an input feature in the classifier, which determines the probabilities that each cluster belongs to a given cosmological model. However, $R_{500}$ remains, in principle, a cosmology-dependent quantity and therefore an \textit{a priori} potential source of cosmology-dependent systematics, although its inclusion in our feature set is both statistically and observationally motivated.

Statistically, our feature-importance analysis, in Fig.~\ref{fig:feature}, shows that $R_{500}$, together with $M_{\mathrm{gas}}$, is among the most informative features for distinguishing between different cosmological models. Observationally, $R_{500}$ defines the aperture within which all integrated X-ray quantities in the \textit{eROSITA} catalogs are measured, making it inseparable from the observables themselves. Unfortunately, the integrated X-ray observables measured inside $R_{500}$---namely $\log M_{\mathrm{gas}}(R_{500})$, $\log L_{\mathrm{gas}}(R_{500})$, and $\log T_{\mathrm{gas}}(R_{500})$--cannot be corrected to account for the cosmology-corrected $R_{500}$, as this would require re-analysis of the underlying surface-brightness and temperature profiles for every cluster. However, previous studies \citep{Vikhlinin_2006, Arnaud_2010, Kravtsov_Borgani_2012, Ghirardini_2019} have demonstrated, in different contexts, that self-similarities in pressure, gas mass, and temperature profiles imply that shifts in $R_{500}$ are expected to have a minor impact on integrated observables.  As a consequence, modest shifts in $R_{500}$ induce only small variations in integrated quantities, and thus the use of catalogued values of $M_{\mathrm{gas}}$, $L_{\mathrm{gas}}$, and $T_{\mathrm{gas}}$ introduces negligible systematics in our analysis. As a consequence, the strongest bias introduced by $R_{500}$, should come from the inclusion of it as a input feature rather than as an aperture for the integrated cluster quantities.  

\begin{figure}
    \hspace{-0.5cm}
    \includegraphics[width=0.5\textwidth]{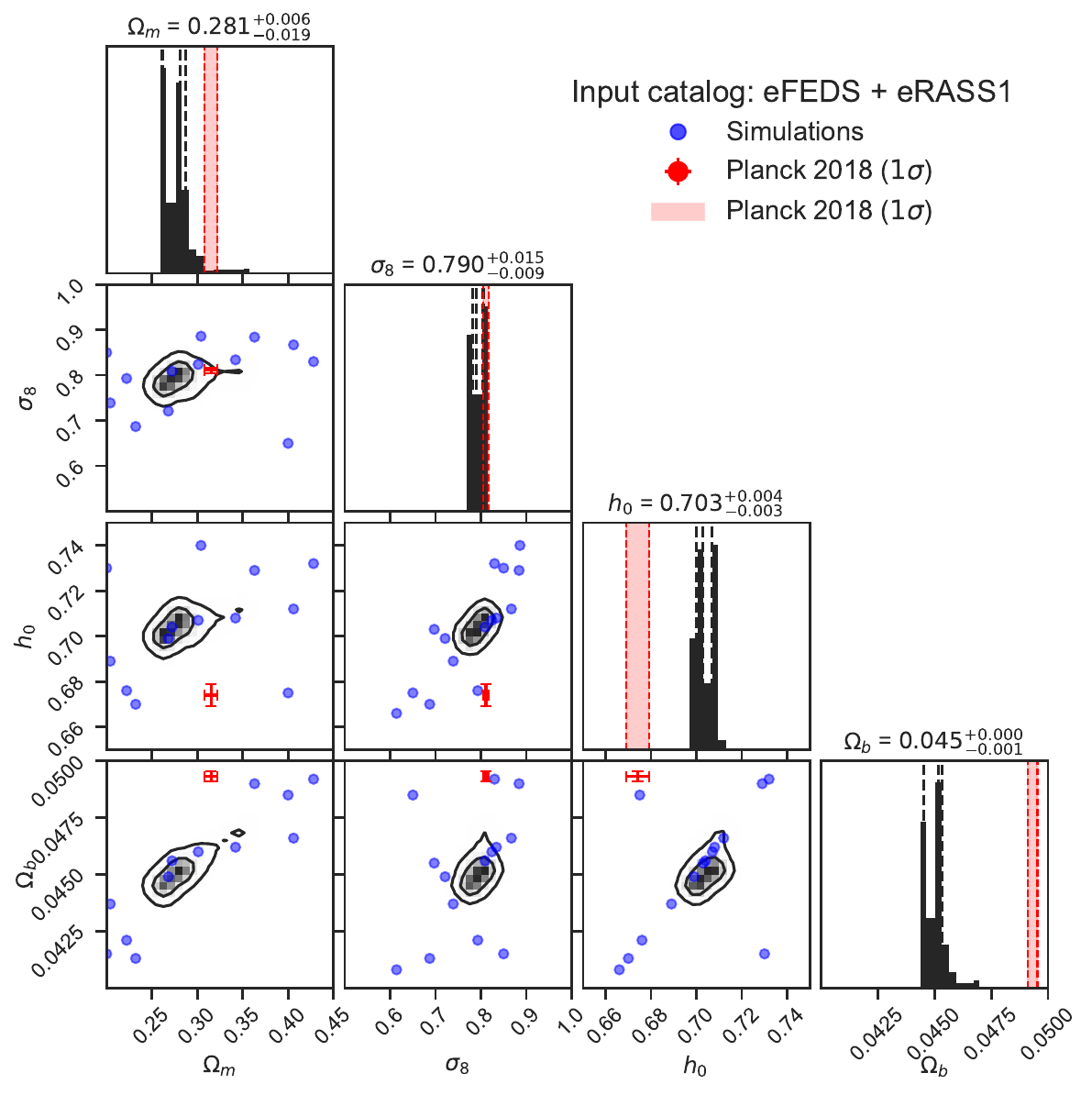}
    \caption{{The cosmology inference results with no $R_{500}$ as the input feature. Planck parameters are shown as red points or shaded regions as in Fig.~\ref{fig:constraints}}.}
    \label{fig: no_R500}
\end{figure}

To test this, we repeated the full cosmological inference after removing $R_{500}$ from the input feature set. The resulting posteriors show no significant shifts in any of the inferred cosmological parameters, although the confidence contours show irregular shapes due to local minima, as shown in {Fig.~\ref{fig: no_R500}}. This shows that $R_{500}$ does not bias inference, while it adds independent information that enhances the significance of our constraints (see also Fig.~\ref{fig: Cosmological_vs_test}).

\section{Conclusions}
\label{sec: conlusions}
In this paper, we have presented the first application of a simulation-based machine-learning inference to real galaxy cluster data from the eROSITA survey. A Random Forest classifier has been trained on multi-cosmology hydrodynamical simulations from \textit{Magneticum} and used to directly connect observable X-ray properties of galaxy clusters from eROSITA, to cosmological parameters, without relying on an explicit mass calibration through scaling relations.

Using the combined eFEDS and eRASS1 cluster samples, we obtained constraints on the main cosmological parameters:
$$
\Omega_m = 0.30^{+0.03}_{-0.02}, ~
\sigma_8 = 0.81 \pm 0.01, ~
h_0 = 0.710 \pm 0.004.
$$
The inferred values of $\Omega_m$ and $\sigma_8$ are consistent with recent large-scale structure and weak lensing measurements and do not show significant tension with CMB constraints. The inferred Hubble constant lies between early-Universe (CMB) and late-Universe determinations and is closer to TRGB-based measurements. We also find an offset in $\Omega_b$, however, our validation tests indicate that this is most likely driven by the limited coverage of the cosmological parameter space in the simulations rather than by baryonic feedback modeling.

A key aspect of this analysis is that cosmological information is extracted directly from observable cluster quantities (gas luminosity, $L_{\rm gas}$, mass, $M_{\rm gas}$, temperature, $T_{\rm gas}$, and radius) in different redshift bins, between $z=0$ and $z=0.8$, rather than from calibrated mass-observable, e.g. the mass functions. This provides a complementary approach to traditional cluster cosmology and reduces systematic uncertainties associated with cluster mass calibration.

We carried out a dedicated assessment of systematic uncertainties affecting the inference. 
Using independent mock catalogues extracted from the simulations, we verified that the 
method recovers the input cosmology without significant bias within the statistical 
uncertainties. We then investigated observational systematics by forward-modeling the 
survey selection function and completeness, finding the inferred parameters to be stable 
against variations in the selection thresholds and sample incompleteness.

To test the robustness of the information content, we repeated the inference after removing $R_{500}$ from the input feature set. The resulting constraints remain consistent, showing that the cosmological signal is not driven by an implicit mass proxy but by the joint distribution of X-ray observables (luminosity, temperature, and gas mass). Feature importance tests further confirm that no single observable dominates the constraints.

Finally, we explored modeling systematics associated with the simulations, including the treatment of baryonic physics and the finite extent of the cosmological parameter space. 
While the recovered cosmology is stable across the various tests, boundary effects in the simulation grid introduce a mild bias, mainly affecting $\Omega_b$. Overall, the systematic uncertainties remain subdominant to the statistical errors for the present sample. A full description of the validation and robustness tests is provided in Appendix \ref{sec: more tests}.

This work demonstrates that ML inference based on hydrodynamical simulations can be applied to real cluster observations to obtain competitive cosmological constraints. Future progress will require broader multi-cosmology simulation suites and larger cluster samples from upcoming eROSITA data releases and next-generation surveys. In this context, simulation-based inference represents a promising complementary route for precision cosmology alongside traditional likelihood-based approaches.

\section{Software}
The code/software used in this article included {\it Numpy}   \citep{harris2020array}, {\it corner}   \citep{corner}, {\it pandas}   \citep{reback2020pandas}, {\it Sklearn}   \citep{sklearn_api}, {\it Matplotlib}   \citep{Hunter:2007}, {\it seaborn}   \citep{Waskom2021}, {\it GetDist}   \citep{Lewis:2019xzd}, {\it Astropy}   \citep{2022ApJ...935..167A}.

\bibliography{ref}
\bibliographystyle{aa}

\appendix

\section{AGN Feedback Impact Analysis}
\label{sec: agn feedback}
\begin{figure}
    \centering
    \includegraphics[width=0.45\textwidth]{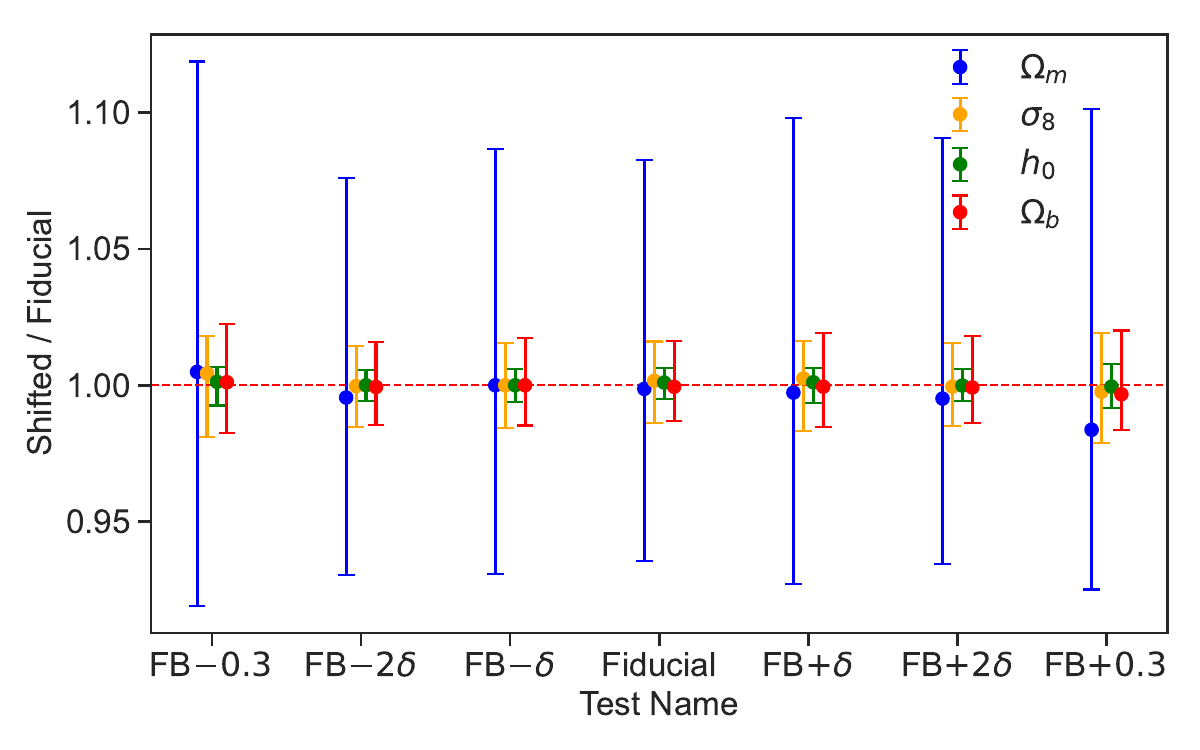}
    \caption{The cosmological parameters and their $1\sigma$ uncertainties varied for different AGN feedback(FB) efficiencies. The y-axis shows the cosmological parameters predicted by models trained on feature-shifted simulation data, normalized by those predicted by the model trained on simulations with the original AGN feedback efficiency. In the x-axis labels, the $\pm \delta$ denotes simulations with physically different AGN feedback settings (from the C8 set); the subscripts $\pm 2\delta$ denote artificially shifting by a factor of two of $\delta$; and the label without a subscript, RF, refers to the original AGN feedback efficiency of 0.15 used in 15 simulations.}
    \label{fig: Cosmological_Parameters_vs_delta}
\end{figure}

The simulation suite used in this work makes use of a single feedback model for all simulations. This can be a limitation in general, as the feedback is expected to have an impact on the efficiency of the star formation and the thermal mechanisms that determine the gas mass or gas temperature associated to a cluster, at a fixed cosmology. In order to investigate the impact of the baryonic physics \citep{Biffi2016, 2018MNRAS.475..648P}, we have checked how variation of the feedback model impacts cosmological inferences. 
We have examined three variants of the AGN feedback efficiencies (0.10, 0.15, and 0.20; the fraction of the back hole accretion energy thermally coupled to the surrounding gas, \citealt{2014MNRAS.442.2304H}) for the Magneticum simulation C8. We determined feature shifts in observables (gas mass, temperature, luminosity) induced by varying feedback strength and recalibrated simulations accordingly.
We calculated the shift of cluster features when changing the AGN feedback and then applied it to the other cosmology simulations. Those features-shifted are calculated based on the median of C8 in different total mass bins (20 bins in the range of $10^{13} - 10^{15} \ M_{\odot}$), which is defined as:
\begin{align}
   & \boldsymbol \delta_- = \text{Median}(\boldsymbol x_{0.15}) - \text{Median}(\boldsymbol x_{0.10}), \\
   & \boldsymbol \delta_+ = \text{Median}(\boldsymbol x_{0.15}) - \text{Median}(\boldsymbol x_{0.20}), \\
   & \boldsymbol x_\pm = \boldsymbol x + \boldsymbol \delta_\pm,
\end{align}
where the index corresponds to the AGN feedback efficiency.
Then we use those new features-shifted simulated data to train a new RF model and predict the cosmological parameters with the observed data as input. Then we proportionally increase those shifting to mimic the more significant impact of AGN feedback.
This approach enabled us to assess the robustness of cosmological constraints against AGN feedback variations. Fig.~\ref{fig: Cosmological_Parameters_vs_delta} shows the impact of different feedback powers.
Here, we see that feedback variations produce no dramatic impact on the cosmological parameters inferences and should not represent a major systematic, except for very extreme feedback models, e.g., the one marked with +0.3, meaning that the scaling relations have been scaled by a factor of almost two with respect to the reference feedback adopted for our analysis. 
This also suggests that the discrepancy found with Planck about $\Omega_b$ might not reside in the effect of feedback. Interestingly, a similar discrepancy is seen also in the combined results of MM-XXL C1 and KiDS, as shown in Fig.~\ref{fig: Comparison_Constrain}, which copes with a consistency in the $h_0-\Omega_m$ plot in the middle panel, both leaving the ground to explanation beyond the $\Lambda$CDM, that might also account for the variation of the cosmological parameters with redshift shown in Fig.~\ref{fig:Parameters_vs_z}).

\section{Testing Different Cosmology Priors, Noise, Classifier configuration, Training size, Redshift Range}
\label{sec: more tests}
\begin{figure*}
    \centering
    \includegraphics[width=0.97\textwidth]{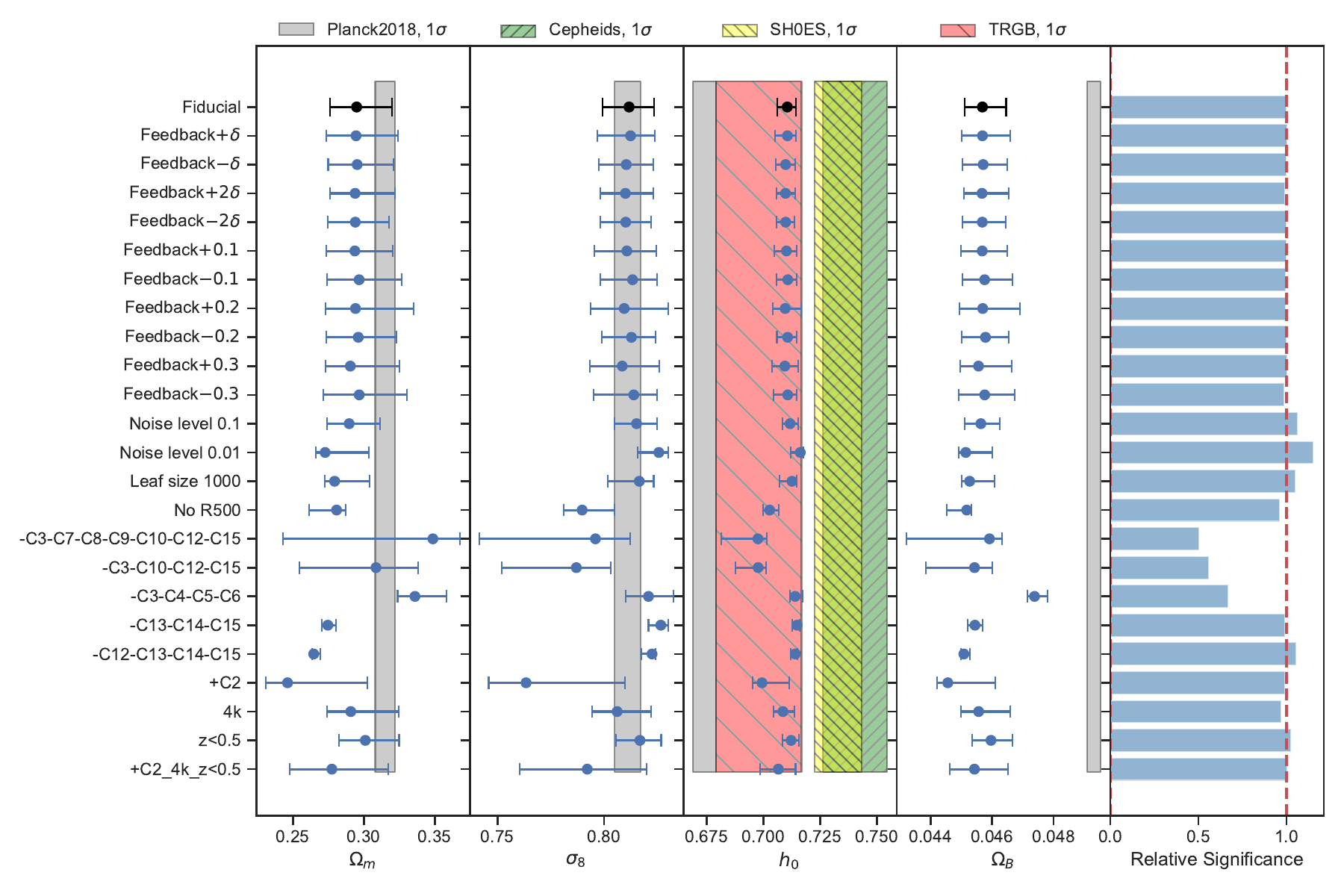}
    \caption{The predicted value and error of 4 cosmological parameters using different configurations. The Planck 2018 results, Hubble constant from Cepheids \citep{2019ApJ...876...85R}, SH0ES \citep{2022ApJ...934L...7R}, and Tip of the Red Giant Branch \citep[TRGB]{2019ApJ...882...34F}.}
    \label{fig: Cosmological_vs_test}
\end{figure*}

To assess the robustness of the RF approach and evaluate the impact on the final cosmological inferences of the parameter space covered by the simulations, feedback recipes, noise adopted for reproducing observational conditions in the training set, and other assumptions behind our analysis, we train and test the network using various combinations of simulations as training sets, different model hyperparameters, noise levels, training size, redshift range, and feedback efficiencies. The prediction and 1-$\sigma$ error are shown in Fig.~\ref{fig: Cosmological_vs_test}, where the labels of the different models are mostly self-explanatory. In this Figure, ``Fiducial" represents the standard results used in the main text, from the set of 13 cosmologies with large enough catalogs ($\sim 10^4$ clusters) to perform a robust training. Then, from top to bottom, we show the test with different feedback recipes as reported in the previous section, noise levels (relative errors on the features), smoothness of the classifier (leaf size), and the test excluding the $R500$ feature, all showing almost no deviation from the ``Fiducial'' model.

We then move to the tests on the priors.
In the test `-C3-C7-C8-C9-C10-C12-C15", we remove C3, C7, C8, C9, C10, C12, and C15 to ensure that the prior of $h_0$ is uniformly distributed around the Planck2018 value, to examine the impact of this prior on the $h_0$ estimation. It shows that $h_0$ shifts toward the Planck value, but at a lower significance of the overall corresponding model (see discussion below).
In the ``-C3-C10-C12-C15" test, we remove C3, C10, C12, and C15 simulations from the Fiducial training set, to exclude a range of high $\sigma_8$ values (also $h_0>0.72$) and check if this strongly affected the confidence contour elongation; in the -C12-C13-C14-C15 or -C13-C14-C15 we tested the impact of excluding the high $\Omega_m$ values ($\Omega_m>0.35$), showing a higher impact over the overall inferences, indicating that the $\Omega_m$ priors are the ones mostly affecting the prediction power. This is confirmed by the -C3-C4-C5-C6, which produces a biased inference toward the high $\Omega_m$ solution. To explore more the impact of the priors, in particualar due to the parameter space coverage, in the ``+C2" test, we add the C2 simulation to the training set, to maximize the parameter volume with yet a reasonable number of clusters for training (C1 is too undersampled) and see a shift in the overall inferences toward low $\Omega_m$ and $\sigma_8$ values, which might come from an unbalanced training sample (C2 by construction has less clusters). In the ``4k" test, we only used $4 \times 10^3$ clusters as training data, to check if a smaller training sample would impact the overall ``Fiducial'' results, {ensuring a balanced number to accommodate the later C2 included test}; in the ``$z<0.5$" test, we also test the case all redshifts are smaller than 0.5 to check the prediction from a higher completeness sample: in both cases there is no substantial change with respect the ``Fiducial'' inferences. Finally, we have combined a ``4k" and ``$z<0.5$" plus the addition of C2 and checked that also the inclusion of a wider parameter volume does not produce problematic deviations with respect to the ``Fiducial'' case, albeit the smaller training sample strongly impacts the confidence of the constraints, producing wider error bars. 

Despite the RF, by construction, always producing a final guess for the dataset it is fed with, we need to eventually check the intrinsic significance of the solution. A first intuitive way to see this is to look at the confusion matrix. Let us assume, for instance, that we have a sample of clusters from cosmology C5, and the prior includes only cosmologies from C6 to C15. According to the confusion matrix, there is still enough overlap with C6 and C7 to place a prediction around the parameters of these cosmologies, but the true probability would be larger if the full parameter space around the true solution is covered, hence leaving the true significance of the model lower than the wider parameter case. This significance should come from the overall superposability of the observed sample and the one reconstructed from the weighted contribution of all cosmologies produced by the RF model. In the example of the C5 above, from the confusion matrix, this superposability would be minimal if the C5 and below simulations are excluded, hence producing a lower significance.    
{In the right panel of Fig. \ref{fig: Cosmological_vs_test}, we define a ``significance" value for each test by the chi-square test for independence between the reconstructed dataset from the simulations and the observation set. We reconstruct a set by randomly sampling clusters from the simulations, proportional to their average predicted probabilities from the classifier in each test. We sample the same number of simulated clusters and then divide both sets (reconstructed bins, $R_i$, and observed bins $O_i$) into small bins. The bins grid is drawn by setting equal sampling with 10 bins pre $(R_{500}, M_{\rm gas}, L_{\rm gas}, T_{\rm gas})$'s corresponding range. Then the reduced chi-square was calculated as follows:
\begin{align}
    \chi^2 = \sum_{i} \frac{(O_i - R_i)^2}{(10^4 -1)}.
\end{align}
Assuming we have the test $j$ and the corresponding $\chi^2_j$, we define a maximum chi-square, $\chi^2_{\rm max}$, which is the chi-square for the lowest probability simulation, and the reference chi-square, $\chi^2_{\rm ref}$, corresponding to the Fiducial result, then the significance is defined as follows:
\begin{align}
    \text{Significance} = \frac{\chi^2_{\rm max} - \chi^2_j}{\chi^2_{\rm max} - \chi^2_{\rm ref}}.
\end{align}
}
From this quantity we can see that, relatively to the ``Fiducial'' case, models that strongly deviate from the fiducial inferences by more than 1$\sigma$ in more than one parameter also show lower significance. We can also see that the absence of important features, such as $R_{500}$, results in lower significance.

All these tests demonstrate that the network is resilient to the variation of the parameter space coverage, if the full domain remains the same, while it loses accuracy if the parameter space does not cover the domain of the original solution (see e.g., -C3-C4-C5-C6 and -C3-C10-C12-C15 cases).

This fully supports the need to secure the necessary multi-cosmology simulations needed to cover the widest range of cosmological parameters in order to further test the impact of priors. This is especially important to clarify whether the solutions we are exploring above are a reflection to ``local'' vs. ``global'' likelihood maxima. Nonetheless, the current parameter volume is already wide enough to identify credible solutions, although further tests will allow us to consolidate the confidence contours of the ML predictions. It goes without saying that the inclusion of updated feedback recipes will allow us to realistically assess the impact of the feedback on the cosmology inferences, in particular, to have a better insight into the $\Omega_b$ tension. 

\end{document}